\documentclass[12pt,preprint]{aastex}

\newcommand{\rns}{R_{\rm ns}}
\newcommand{\rlc}{R_{\rm lc}}
\newcommand{\zobs}{\zeta_{\rm obs}}
\newcommand{\thb}{\theta_{\rm b}}
\newcommand{\thm}{\theta_{\rm m}}
\newcommand{\thov}{\theta^r_{\rm ov}}
\newcommand{\wi}{W}
\newcommand{\rgeo}{r_{\rm geo}}
\newcommand{\rdel}{r_{\rm del}}
\newcommand{\vbret}{\vec B_{\rm ret}}
\newcommand{\vbstat}{\vec B_{\rm st}}
\newcommand{\bstatx}{B_{{\rm st,} x}}
\newcommand{\bstaty}{B_{{\rm st,} y}}

\newcommand{\bretx}{B_{{\rm ret,} x}}
\newcommand{\brety}{B_{{\rm ret,} y}}
\newcommand{\bretz}{B_{{\rm ret,} z}}
\newcommand{\bretr}{B_{{\rm ret,} r}}
\newcommand{\bretp}{B_{{\rm ret,} \phi}}
\newcommand{\brett}{B_{{\rm ret,} \theta}}
\newcommand{\bret}{B_{\rm ret}}
\newcommand{\bstat}{B_{\rm st}}
\newcommand{\vdb}{\Delta \vec B}
\newcommand{\rn}{r_{\rm n}}
\newcommand{\pl}{P_{\rm l}}
\newcommand{\pt}{P_{\rm t}}
\newcommand\kl{\kappa_{\rm l}}
\newcommand\kt{\kappa_{\rm t}}
\newcommand\xl{x_{\rm l}}
\newcommand\yl{y_{\rm l}}
\newcommand\zl{z_{\rm l}}
\newcommand\xt{x_{\rm t}}
\newcommand\yt{y_{\rm t}}
\newcommand\zt{z_{\rm t}}
\newcommand\om{(\vec \Omega, \vec \mu)}
\newcommand\dbx{\Delta B_x}
\newcommand\dby{\Delta B_y}
\newcommand\dbz{\Delta B_z}
\newcommand\dbr{\Delta B_r}
\newcommand\dbp{\Delta B_\phi}
\newcommand\dbt{\Delta B_\theta}
\newcommand\phl{\phi_{\rm l}}
\newcommand\pht{\phi_{\rm t}}
\newcommand\dphasym{\Delta\phi_{\rm l-t}}
\newcommand\dphasymtwo{\Delta\phi_{\rm s-r}}

\newcommand\displ{\delta_{\rm ov}}

\newcommand\dphnew{\Delta\Phi}
\newcommand\dphpa{\Delta\phi_{\rm PA}}
\newcommand\dphp{\Delta\phi_{\rm pf}}
\newcommand\dphov{\Delta\phi_{\rm ov}}
\newcommand\apan{\beta_{\rm x}}

\newcommand\sss{\scriptscriptstyle}
\newcommand\tmin{\tau_{\rm min}}
\newcommand\tmax{\tau_{\rm max}}
\newcommand\rmin{\rn^{\rm min}}
\newcommand\rmax{\rn^{\rm max}}
\newcommand\fmin{F_{\rm min}}
\newcommand\fmax{F_{\rm max}}

\shorttitle{Polarization in Pulsar Models}
\shortauthors{Dyks \& Harding}

\begin{document}

\title{Rotational sweepback of magnetic field lines\\
in geometrical models of pulsar radio emission}

\author{J. Dyks\altaffilmark{1} and Alice K. Harding}
\affil{Laboratory for High Energy Astrophysics,
    Greenbelt, MD 20771, USA}
\email{jinx@milkyway.gsfc.nasa.gov, harding@twinkie.gsfc.nasa.gov}

\altaffiltext{1}{On leave from Nicolaus Copernicus Astronomical Center,
Toru{\'n}, Poland}

\begin{abstract}
We study the rotational distortions of the vacuum dipole magnetic field
in the context of geometrical models of the radio emission from
pulsars. We find that at low altitudes the rotation deflects
the local direction of the magnetic field by at most an angle 
of the order of $\rn^2$, where $\rn = r/\rlc$, $r$ is the radial
distance of the radio emission and $\rlc$ is the light cylinder radius. 
To the lowest (ie.~second) order
in $\rn$,
this distortion is symmetrical
with respect to the plane containing the dipole axis and the rotation
axis ($\om$ plane). The lowest order distortion which is asymmetrical
with respect to the $\om$ plane is third order in $\rn$.
These results confirm the common assumption that the rotational
sweepback has negligible effect on the position angle (PA)
curve. We show, however, 
that the influence of the sweepback on the
outer boundary of the open field line region (open volume) 
is a much larger effect, of the order of $\rn^{1/2}$. The open volume
is shifted backwards with respect to the rotation direction by an angle
$\delta_{\rm ov} \sim 0.2\sin\alpha \rn^{1/2}$ where $\alpha$ is the
dipole inclination with respect to the rotation axis.
The associated phase shift of the pulse profile
$\Delta\phi_{\rm ov} \sim 0.2\rn^{1/2}$ can easily exceed 
the shift due to combined effects of aberration and propagation time
delays ($\approx 2\rn$). This strongly affects 
the misalignment of the center of the PA curve
and the center of the pulse profile, thereby modifying
the delay-radius relation. 
Contrary to intuition, the effect of sweepback dominates over other
effects when emission occurs at low altitudes. 
For $\rn \la 3\cdot 10^{-3}$ the shift becomes negative, ie.~the center
of the position angle curve precedes the profile center. 
With the sweepback effect included, the modified delay-radius relation 
predicts larger emission radii and is in much better agreement with
the other methods of determining $\rn$.
\end{abstract}


\section{Introduction}

There are independent observational arguments which 
imply that the pulsar radio emission occurs in a form of a narrow beam
centered (or roughly centered) on the magnetic dipole axis.
In many cases the position angle (PA) of the observed 
linearly polarized radiation
changes its direction by nearly $180^\circ$ when our line of sight 
crosses the radio beam (eg.~Lyne \& Manchester 1988). 
If associated with the direction of $\vec B$,
this change of PA can be naturally interpreted as a result of our line
of sight passing near the magnetic pole (Radhakrishnan \& Cooke 1969).
Moreover, the width $\rho$ of the radio beam determined for different pulsars
from the observed width of their radio pulse profiles scales with the 
rotation period $P$ as the opening angle of the open field line region,
ie.~$\rho \propto P^{-1/2}$ (Rankin 1990; Rankin 1993). 

It is commonly believed that the emission region 
associated with the beam does not extend beyond
the region of open field lines (hereafter called ``open volume")
which cross the light cylinder of radius $\rlc = c/\Omega$
($c$ is the speed of light and $\vec \Omega= \Omega \hat z$ 
is the angular velocity of
pulsar rotation). The angular size of the open volume 
at (small) radial distance $r$ is equal to 
$\thov \simeq (r/\rlc)^{1/2}$ and the cone formed by tangents to
magnetic field lines at the rim of the open volume has angular radius 
of $\thb \simeq 1.5\thov$.
The radial distance of the emission region has not been established
so far: both a high-altitude emission region 
extending over a small fraction of $\thov$, 
as well as a low-altitude emission region which fills in a much larger 
fraction of $\thov$
may be responsible for the same shape of the radio beam.

The sweepback effect was first investigated in detail by
Shitov (1983) who considered it to explain the observed
dependence of radio luminosity of pulsars as a function of period.
He estimated the magnitude of the rotational distortions of the magnetic
field from the torque responsible for the observed slowing down of
pulsars. He found that at moderate altitudes within the open volume,
``near" the dipole axis,
the direction of the distorted magnetic field deflects from the 
direction of the pure (ie.~static shape) dipole barely by an angle
\begin{equation}
\delta_{\rm sb} \simeq 1.2\left(\frac{r}{\rlc}\right)^3 \sin^2\alpha,
\label{deflection}
\end{equation}
where $\alpha$ is the dipole inclination with respect to the rotation 
axis.
Gil (1983) proposed the sweepback effect to explain why the separation
between the main radio pulse and the interpulse observed in the profile
of PSR B0950$+$08 is significantly different from $180^\circ$.

In 1985 Shitov incorporated the sweepback effect into 
the model of pulsar position angle curves proposed by 
Radhakrishnan \& Cooke (1969)
and showed that the sweepback
results in a lag of the profile center (measured as the midpoint between
the outer edges of the pulse profile) with respect to the center,
or the ``inflection point" of the position angle curve.
Shitov emphasized that the lag of the profile center was a sum of
\emph{two} effects: not only the center of the PA curve is shifted
toward earlier phases (with respect to the nondistorted case) according
to the eq.~(\ref{deflection}), but also the center of the
open volume is displaced backwards, which contributes to the total
effect.

In most of subsequent investigations, however, the sweepback has been
neglected, mainly on the basis of eq.~(\ref{deflection}).
Blaskiewicz et al.~(1991, hereafter BCW91) proposed a relativistic model
of pulsar polarization which took into account two important
effects overlooked by Shitov: the presence of
the corotational acceleration and the aberration effect.
An excellent result of their work was the ``delay-radius" relation,
according to which the center of the PA curve \emph{lags} the profile
center by
\begin{equation}
\Delta\Phi_{\sss BCW} \approx 4\frac{r}{\rlc}\ {\rm rad,}
\label{bcw}
\end{equation}
where $r$ is the radial distance of the radio emission.
With no dependence on viewing geometry parameters (like the dipole
inclination $\alpha$, or the viewing angle $\zobs$ between the rotation
axis and the observer's line of sight), their relation appears to
provide a powerful method of determining $r$.
Equally important, the delay-radius relation depends neither on the
observed width of the pulse profile $\wi$ nor on the separation between
the conal components in the pulse profile.
Therefore, the altitudes of radio emission provided by eq.~(\ref{bcw})
may serve to determine which magnetic field lines are associated with
the outer edge of the profiles and which field lines correspond
to the maxima of conal components (Mitra \& Rankin 2002; Dyks et
al.~2004a). Von Hoensbroech \& Xilouris (1997) used the delay-radius relation
to probe the radius-to-frequency mapping at high radio frequencies.  
Given that the method is based on a measurement of \emph{tiny} shifts
(of magnitude usually being a small fraction of one degree) between the
centers of the PA curve and the profile, it is extremely sensitive
to the assumed geometry of the magnetic field. The latter was taken to
be a dipole of static shape, with no rotational distortions.

Gangadhara \& Gupta (2001) proposed another relativistic method
of estimating radio emission altitudes for pulsars with both core and
conal components. By considering the effects of the aberration and the
propagation time delays they showed that the core component lags in
phase the midpoint between the maxima of conal components, if the core
originates from lower altitudes than the cones, and if
the cones are axially symmetric around the core in the reference frame
corotating with the star (CF). 
Dyks et al.~(2004a) revised their method and
showed that the phase shift between the core component and the pairs
of conal components is equal to
\begin{equation}
\Delta\phi_{\sss DRH} \approx 2\frac{r}{\rlc}\ {\rm rad}
\label{drh}
\end{equation}
which provides another method for determining $r$ without information
about viewing geometry (nor $\wi$).
Dyks et al.~(2004a) used the above relation, and the results of work
by Gupta \& Gangadhara (2003) to calculate $r$ for 6 pulsars with well
defined core-cone systems. 
As in the case of the
delay-radius relation, the above formula holds only for
the magnetic field which is symmetrical about the $\om$ plane
(where $\vec \mu$ is the magnetic moment of the pulsar magnetic field),
at least as long as one associates the assumed symmetry of the core-cone
system with the geometry of the underlying magnetic field. Any asymmetrical
(with respect to the $\om$ plane) distortions of the magnetic field 
would pose a serious problems for the framework of the model leading to
eq.~(\ref{drh}). 
Again,  based on eq.~(\ref{deflection}),
any influence of the sweepback was neglected.

Given that the above-mentioned methods of determining $r$ 
are so sensitive to the assumed 
symmetry of the magnetic field around $\vec \mu$, it is important
to have the symmetry hypothesis well justified.
It is also important to revise this assumption in view of the 
unacceptably low
values of $r$ which are often being derived with the BCW91 method:
as found in BCW91, the ``delay radii" $\rdel$
implied by their method (eq.~\ref{bcw}) 
are often smaller (in some cases by an order of
magnitude --- see fig.~29 in BCW91) than the geometrical radii
$\rgeo$ determined with the traditional geometrical method based 
on the measurement
of profile widths (Cordes 1978; Gil \& Kijak 1993; Kijak \& Gil 2002).
This poses a real problem for the
BCW91 method, because the geometrical radii, in the absence of strong
refraction effects (Lyubarski \& Petrova 1998), should be considered as lower
limits of $r$ (Dyks et al.~2004a). 
Although one could
explain this disagreement in many different ways (eg.~underestimated
theoretical width of the open volume, 
systematically overestimated impact angles and dipole inclinations $\alpha$ 
etc.), we show below that the
rotational distortions of the static shape dipole may account for
a large part of 
the discrepancies between $\rdel$ and $\rgeo$.

Recently Kapoor \& Shukre (2003) considered the aberration effect
\emph{and} the rotational sweepback to investigate the relative locations
of core and cone components in the pulsar magnetosphere. Although included 
in the model, the sweepback is again estimated with the help of
eq.~(\ref{deflection}). Being aware of the limitations of Shitov's
estimate, the authors emphasized the need for derivation of a
more advanced formula
describing the rotational distortions of the magnetosphere.
They noted that a proper derivation ``should make use of at least the
magnetic field given by the full Deutsch solution (Deutsch 1955)". 

Such an estimate based on the Deutsch solution was done by Arendt \&
Eilek (1998), who concluded that the rotation distorts the magnetic
field by a magnitude of the order of $r/\rlc$. Being much larger than
the Shitov's estimate, this distortion would strongly affect results in
BCW91, GG2001, Hibschman \& Arons (2001, hereafter HA2001), 
and Dyks et al.~(2004a). 
On the contrary, HA2001
noted that the leading terms in the difference between 
the Deutsch field and the rigidly rotating static-shape dipole
are of the order of $(r/\rlc)^2$.  
Recently, Mitra \& Li (2004) emphasized that on the theoretical side
there is a great need to develop
and understand the details of the sweepback effect.

In this paper we investigate the rotational
distortions of the pulsar magnetic field assuming the approximation
of the vacuum magnetosphere.
The twofold nature of the sweepback, first noticed by Shitov (1983)
will be highlighted, and limitations in applicability 
of eq.~(\ref{deflection}) will be clarified (Section 2).
The significance of the sweepback for the relativistic model of pulsar
polarization
will appear to be much larger than previously thought, which will have 
serious consequencies for the delay-radius relation (eg.~modification
of eq.~\ref{bcw}, Section 3).

\section{The rotational distortions of the dipolar magnetic field}

We follow previous investigators (Deutsch 1955; Shitov 1983; Barnard 1986;
Romani \& Yadigaroglu 1995, hereafter RY95; Cheng et al.~2000, hereafter
CRZ2000)
in assuming that the magnetic field
surrounding the neutron star (NS)
may be approximated by the vacuum rotating dipole.
As in Barnard (1986), RY95, and CRZ2000, we assume that 
outside the NS the field is the same as of
the star-centered \emph{point} dipole, 
ie.~we neglect the near-surface modifications of the magnetic field
by the conducting sphere of the neutron star, derived by Deutsch
(1955) (cf.~Yadigaroglu 1997). Hereafter, the magnetic field
will be called a ``retarded dipole" and will be denoted by $\vbret$.
In Appendix \ref{appenA} we give the cartesian and the spherical
components of $\vbret$ (eqs.~\ref{bretx} -- \ref{bretx} and \ref{bret1} --
\ref{bret3}, respectively).

We want to estimate how much the rotational sweepback distorts
the magnetic field at low altitudes ($r \ll \rlc$). One measure of this
is the difference between the retarded
magnetic field $\vbret$ and the magnetic field of the
static-shape dipole $\vbstat$. The components of $\vbstat$ can be
calculated with the help of
eqs.~(\ref{bret1} -- \ref{bret3}) 
taken in the limit of $\rn \ll 1$ (ie.~with the ratio $\rn$ set equal 
to zero). We define the difference as:
\begin{equation}
\vdb = \vbret - \vbstat.
\label{db}
\end{equation}
In all formulae we assume that both the retarded dipole and the
static-shape dipole are associated with the same
magnetic moment $\vec \mu$, which at the time $t=0$ 
is in the $(\hat x, \hat z)$ plane
(time $t$ is measured in the
Lorentz frame in which the neutron star's center of mass is at rest). 
Thus, at any instant $\vec \mu_{\rm st} = \vec \mu_{\rm ret} =\vec\mu$,
where
\begin{equation}
\vec \mu = \mu(\sin\alpha\cos\Omega t\ \hat x + 
\sin\alpha\sin\Omega t\ \hat y + \cos\alpha\ \hat z).
\label{mu}
\end{equation}
In the CF the components of $\vbret$ at any point which
corotates with the magnetosphere do not depend on $t$
(cf.~eq.~\ref{rotation}).  
Therefore, one is allowed to choose any convenient value.
We take $t=0$ ($\vec \mu$ in the $(\hat x, \hat z)$ plane) 
and constrain our discussion to the half of the magnetosphere
with positive values of $x$. The positive values of the $y$ coordinate 
then correspond to the leading part of the
magnetosphere and the negative $y$ correspond to the trailing part.
In cartesian coordinates and for $t=0$ the difference is:
\begin{eqnarray}
\dbx & = & \frac{\mu}{r^5}\sin\alpha \left[
\frac{1}{2}\left(x^2 + r^2\right)\rn^2 + O(\rn^4)\right] 
\label{dbx} \\
\dby & = & \frac{\mu}{r^5}\sin\alpha \left[
\frac{1}{2}\ x y\ \rn^2 - \frac{2}{3}\ r^2 \rn^3 + O(\rn^4)\right] 
\label{dby} \\
\dbz & = & \frac{\mu}{r^5}\sin\alpha \left[
\frac{1}{2}\ x z\ \rn^2 + O(\rn^4)\right].
\label{dbz}
\end{eqnarray}
In agreement with the remark of Hibschman \& Arons (2001), the leading
terms are second order in $\rn$. The second order terms of $\dbx$
and of $\dbz$ do not depend on $y$. The second order term of $\dby$
is odd function of $y$ ($\dby(y) = -\dby(-y) + O(\rn^3)$).
These features are important because the symmetry 
of any vector field $\vec B=(B_x, B_y, B_z)$ 
with respect to the $\om$ plane
requires the following relations to be satisfied: $B_x(x, y, z) = B_x(x,
-y, z)$, $B_y(x, y, z) = -B_y(x, -y, z)$, and $B_z(x, y, z) = B_z(x,-y, z)$
(ie.~the $B_x$ and $B_z$ components must be even, and the $B_y$ component must
be odd in $y$).

The angle $\kappa$ between $\vbret$ and $\vbstat$, to the order
$\rn^3$, is given by:
\begin{equation}
\kappa \approx \left(f_1 \rn^4 + f_2 \rn^5 + f_3
\rn^6\right)^{1/2}
\la \rn^2,
\label{kappa}
\end{equation}
where the functions
$f_1$, $f_2$, and $f_3$, given in Appendix \ref{appenA}, depend 
on $x$, $y$, $z$, and on the inclination angle $\alpha$ 
(but not on $\rn$) and in general have magnitude of the order of 1,
except from special locations in the magnetosphere which we discuss below.

Thus, the rotation causes the magnetic field to deviate from $\vbstat$
by an angle which at most is second order in $\rn$.
Along the magnetic dipole axis\footnote{By the ``magnetic dipole axis"
we understand the \emph{straight} line containing the magnetic moment $\vec
\mu$. In the case of the retarded dipole, the axis cannot be associated 
with any magnetic field line.}
of an orthogonal rotator, however,
(ie.~for $\alpha = 90^\circ$ and $(x,y,z) = (r,0,0)$), one obtains
$f_1=0$, $f_2=0$ and $f_3 = 9^{-1}$ so that $\kappa$ is third order in
$\rn$:
\begin{equation}
\kappa = \frac{1}{3}\ \rn^3
\label{kapax}
\end{equation}
in partial agreement with the estimate of Shitov (1983). 
Beyond the orthogonal dipole axis, however, as well as at the dipole axis
of non-orthogonal rotator, $f_1 \ne 0$ and $\kappa$ may be much larger.
On the $\om$ plane $f_2=0$ (because $f_2 \propto y$,
cf.~eq.~\ref{smallefs}).
Beyond the $\om$ plane ($f_1 \ne 0$, $f_2 \ne 0$),
the first two terms in eq.~(\ref{kappa}) dominate and give:
\begin{equation}
\kappa \simeq f_1^{1/2} \rn^2 + \frac{f_2}{2 f_1^{1/2}} \rn^3.
\label{kappa2}
\end{equation}

\begin{figure}
\epsscale{0.7}
\plotone{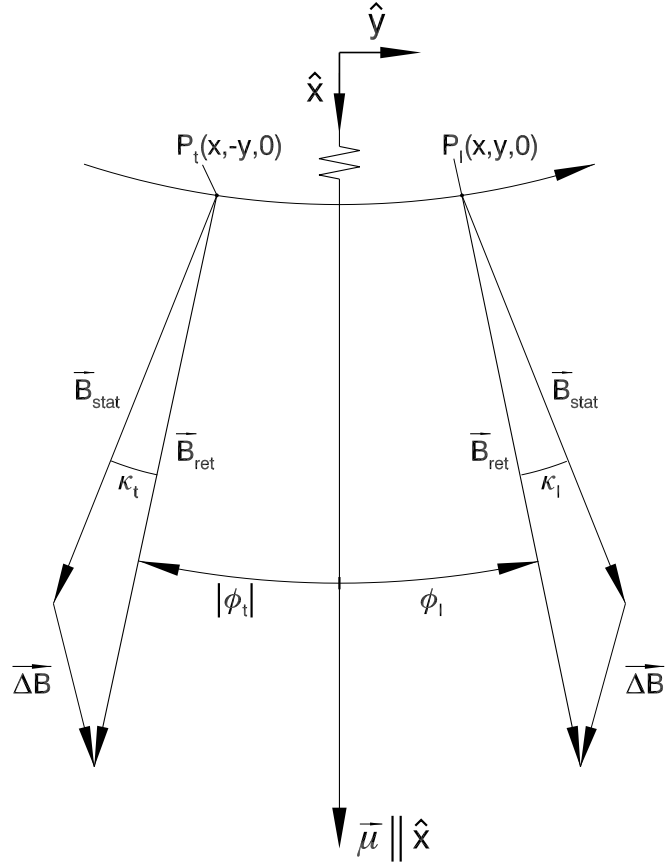}
\caption{The influence of rotation on the local direction of the magnetic
field. The vectors of $\vbret$ and $\vbstat$ are shown for two points
in the equatorial plane of the orthogonal rotator ($\alpha=90^\circ$).
The points are located symmetrically with respect to the $\om$ plane
(the plane is orthogonal to the page and contains $\hat x$).
The rotation is to the right and it is assumed that $x \gg y$ (region
near the dipole axis) and $x \ll \rlc$.
With accuracy of $\rn^2$, the retarded field $\vbret$ is symmetrical 
with respect to the $\om$ plane, ie.~$\kl=\kt$ and
$|\pht|=\phl$. More precisely, $\kl =\kt + O(\rn^3)$ and $\pht = -\phl
- O(\rn^3)$, ie.~$\kt < \kl$ and $|\pht| > \phl$.
\label{distor}}
\end{figure}

Let us estimate the angles $\kl$ and $\kt$
for two points $\pl(\xl,\yl,\zl)$ and $\pt(\xt, \yt, \zt)$
located symmetrically on both sides of the $\om$ plane.
Let us consider the particular case in which the points lie 
in the plane of rotational equator, close to the rim of the open volume 
of orthogonal rotator, ie.~$\xl=\xt=r(1-\rn)^{1/2}$, $\yl=-\yt=r\rn^{1/2}$,
and $\zl=\zt=0$, with the positive value of the $y$
coordinate corresponding to the point $\pl$ on the leading side of the
open volume, and the negative $y$ for the trailing point $\pt$
(cf.~Fig.~\ref{distor}). 
Then eq.~(\ref{smallefs}) gives $f_1 = 4^{-1}\rn$, $f_2 = \pm 3^{-1}
\rn^{1/2}$, which results in 
$\kl \approx 2^{-1}\rn^{5/2} + 3^{-1}\rn^3$ and 
$\kt \approx 2^{-1}\rn^{5/2} - 3^{-1}\rn^3$.
Both angles are considerably larger than the distortion of the magnetic axis
given by eq.~(\ref{kapax}). 

Eqs.~(\ref{dbx} -- \ref{dbz}) imply, however, that up to the second
order in $\rn$, the rotational distortion of $\vbstat$ \emph{is
symmetrical} with respect to the $\om$ plane (because $\dbx$ is even, 
and the leading term of $\dby$ is odd in $y$).  
This (approximate) symmetry, shown in
Fig.~\ref{distor}, implies that \emph{beyond the dipole axis of the
orthogonal rotator}
the angle $\kappa$ between $\vbret$ and
$\vbstat$ provides no information about the magnitude of
the asymmetry of $\vbret$ with respect to the $\om$ plane.

To estimate the asymmetry for points located beyond the $\om$ plane, 
one must therefore use the difference between
azimuths of $\vbret$ at the points $\pl$ and $\pt$:
\begin{equation}
\dphasym \equiv |\pht| - \phl \approx \left|\frac{\brety(\pt)}
{\bretx(\pt)}\right|
- \frac{\brety(\pl)}{\bretx(\pl)} \approx \frac{4}{3}\frac{\mu}{r^3}\ 
\frac{\sin\alpha} 
{\bstatx}\ \rn^3,
\label{dphasym}
\end{equation}
where $\phl$ is the azimuth of $\vbret$ at the point $\pl(\xl, \yl, \zl)$,
$\pht$ is the azimuth of $\vbret$ at $\pt(\xl, -\yl, \zl)$
(cf.~Fig.~\ref{distor}), and 
\begin{equation}
\bstatx = \frac{\mu}{r^3}\left[
3\frac{x}{r}\frac{z}{r}\cos\alpha + \left(3\frac{x^2}{r^2} - 1\right)
\sin\alpha\right].
\label{bstatx}
\end{equation}
Equation~(\ref{dphasym}) clearly demonstrates that $\dphasym$ is third order
in $\rn$, ie.~the rotation induces the asymmetry of $\vbret$ with
respect to the $\om$ plane with magnitude of the order of $\rn^3$.
Equation~(\ref{dphasym}) is not useful
in the immediate vicinity of the $\om$ plane (nor at the plane itself),
because $\phl$ changes sign to negative (ie.~$\vbret$ is parallel to the
$\om$ plane) for locations with the tiny azimuth
\begin{equation}
\phi_\parallel \approx \frac{y}{x} \approx \frac{2}{9}\ \rn^3 \left(\frac{x}{r}\frac{z}{r}
\frac{1}{\tan\alpha} + \frac{x^2}{r^2}\right)^{-1} \simeq
\frac{2}{9}\ \rn^3
\label{phizero}
\end{equation}
(ie.~within a narrow region on the leading side of the $\om$ plane
$\brety < 0$ and eq.~(\ref{dphasym}) gives a sum rather than a
difference of azimuths).
To estimate the asymmetry on the $\om$ plane
one can use the
difference of azimuths of $\vbstat$ and $\vbret$ at a given (the same) point:
\begin{equation}
\dphasymtwo \equiv \phi_{\rm st} - \phi_{\rm ret} \approx
\frac{\bstaty}{\bstatx}
- \frac{\brety}{\bretx} 
\approx f_4 \rn^2 + \frac{2}{3}
\frac{\mu}{r^3} \frac{\sin\alpha}{\bstatx}\ \rn^3,
\label{dphasym2}
\end{equation}
where the function $f_4(x, y, z, \alpha)$ is given in Appendix
\ref{appenA} (eq.~\ref{ef4}).
Beyond the $\om$ plane the first term in this equation dominates
(ie.~$\dphasymtwo \simeq f_4 \rn^2$) but it is symmetrical with respect 
to the $\om$ plane (ie.~odd in $y$, $f_4 \propto y$). 
Therefore, just like
$\kappa$ given by eq.~(\ref{kappa}), $\dphasymtwo$ provides no estimate
of the rotational asymmetry there. 
On the $\om$ plane $f_4=0$
and $\dphasymtwo$ does measure the asymmetry which is of the order of $\rn^3$.
Along the dipole axis $\bstatx = 2\mu\sin\alpha/r^3$ so that 
$\dphasymtwo \simeq 3^{-1} \rn^3$, independent of $\alpha$. 

Eqs.~(\ref{kappa}), (\ref{dphasym}), and (\ref{dphasym2}) 
can be summarized as follows:
the rotation changes the components and the direction 
of the dipolar magnetic field by $\sim \rn^2$. To the order of $\rn^2$
this change, however,
is symmetrical with respect to the $\om$ plane. The \emph{asymmetrical} 
change of the magnetic field direction has much smaller magnitude of 
the order of $\rn^3$ and is given by the second term of
eq.~(\ref{dphasym2}).

An immediate consequence of eq.~(\ref{kappa})
is that with accuracy of the order of $\rn$, 
the rotation does not affect the shape of 
the position angle curve which depends on the
\emph{local} direction of $\vbret$. 
This allows one to neglect the influence of the sweepback 
\emph{on the position angle curve} as long as only the first order effects
in $\rn$ are considered. Given the tiny magnitude of the rotational
asymmetry as defined by eq.~(\ref{dphasym}), investigators often neglect
its influence on the shape of the pulse profile as well (eg.~BCW91, GG2001,
HA2001). As we show below, this is not justified.

\subsection{The rotational distortions of the open field line region}

In the method of BCW91, the center of the pulse profile is most
efficiently measured as a midpoint between the \emph{outer edges} of the
profile. Therefore, the method is based on the assumption
that the \emph{outer boundary} of the open volume
is symmetrical with respect to the $\om$ plane.
Due to the complexity of the magnetic field lines in the retarded case,
we determine the outer boundary of the open volume numerically, by
finding the magnetic field lines which are tangent to the light
cylinder. The method is described in detail in Dyks et al.~(2004b).

\begin{figure}
\epsscale{1.0}
\plotone{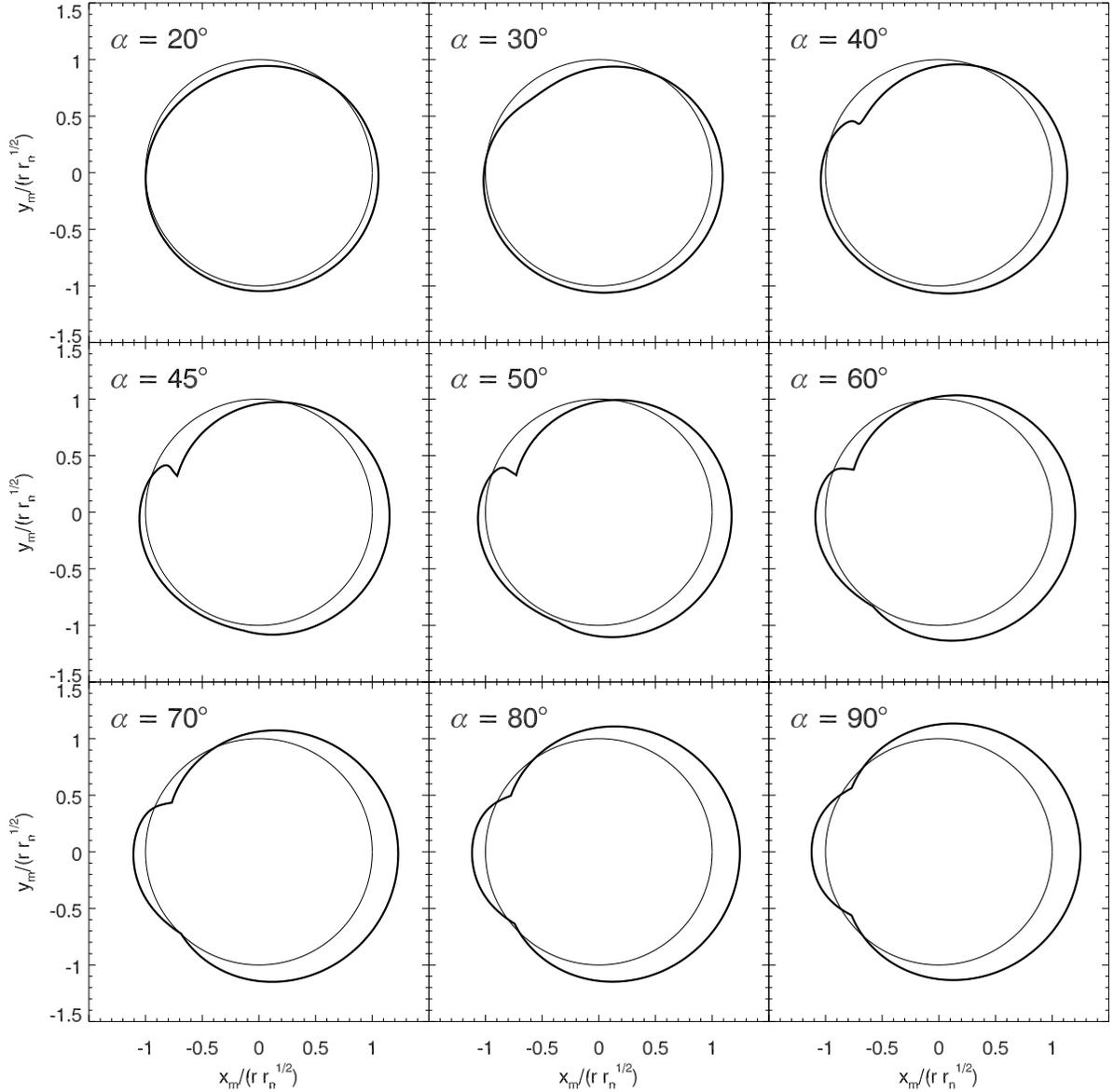}
\caption{Crossections of the open volume with the star-centered 
sphere of radius $r \ll \rlc$ calculated for the retarded dipole field
$\vbret$ (thick solid). The circles have radius of $r\rn^{1/2}$ and are
centered on the magnetic moment $\vec \mu$ which protrudes
perpendicularly from the page at the center of each panel 
($(x_{\rm m}, y_{\rm m}) = (0,0)$).
The magnetic field $\vbret$ which permeates each of the panels
is nearly the same as that of the static-shape dipole with the axis
parallel to $\vec \mu$ (ie.~it protrudes from the $(0,0)$ point too). 
Rotation is to the left. The backward
displacement of the retarded contours with respect to the $(0,0)$ points
(or circles) results in the shift of the center of the pulse profile
toward later phases. The shape of the contours does not depend
on $r$ and $P$ as long as $r \ll \rlc$ ($P = 1$ s, and $\rn=0.01$ was assumed
in the figure).
Their size scales as $r\rn^{1/2}$.
\label{caps}}
\end{figure}

Thick solid lines in Fig.~\ref{caps} present the transverse shape of the
open volume at low altitudes ($\rn \ll 1$), calculated for the retarded
magnetic field $\vbret$.
More precisely, they represent the crossection of 
the outer boundary of the
open volume with a sphere of radius $r=0.01\rlc$
centered at the neutron star. Different panels correspond to different
dipole inclinations $\alpha$. The magnetic moment 
$\vec \mu$ in all panels emerges perpendicularly from the page at the point
$(x_{\rm m}, y_{\rm m}) = (0,0)$. 
The thin circles have radius equal to
$r\thov=r\rn^{1/2}$, and are centered at the $(0,0)$ point to 
guide the eye in assessing the asymmetry of the open volume around $\vec \mu$. 
Given the small difference between the local direction of $\vbret$
and $\vbstat$ (eq.~\ref{kappa}), one may regard each panel to be
permeated by the magnetic field of the \emph{static} dipole with 
the \emph{straight} magnetic field line emerging from the $(0,0)$ 
point toward the reader. 
The field is symmetric relative to the $\om$ plane, ie.~with respect to
the vertical line of $x_{\rm m} = 0$ (to be imagined in
each panel of Fig.~\ref{caps}).
In the course of corotation, the contours of the open volume 
outer boundary move to the left in Fig.~\ref{caps}, ie.~$x_{\rm m} < 0$ 
correspond to the leading,
and $x_{\rm m} >0$ to the trailing side. 
An observer's line of sight cuts the
contours horizontally, moving left to right.\footnote{Note that
our Fig.~\ref{caps}, calculated for $t=0$ in eqs.~(\ref{bretx}
-- \ref{bretz}), 
differs significantly from analogous figures shown in 
Arendt \& Eilek (1998) and in CRZ2000.
The location of the \emph{retarded} polar caps in their figures corresponds to
the magnetic moment $\vec \mu$ rotated by the angle $\Omega \rns/c$ 
with respect to the $(\hat x,\hat z)$ plane
(ie.~they assume $t=\rns/c$ in eqs.~\ref{bretx} --\ref{bretz}).
At the same time, however, they assume $t=0$ 
($\vec \mu$ in $(\hat x,\hat z)$ plane)
to position the polar caps for the \emph{static} case. 
Therefore, their figures do not
inform us what is the relative position of the static and retarded
caps in phase - only the caps'
shapes can be compared. (To enable this, their retarded caps would have to
be derotated by $\Omega \rns/c$ with respect to the static caps.)
Also, note that the components of the
magnetic field for the static-shape dipole given in CRZ2000 (eqs.~A1--A3
therein) are for $t=0$ whereas the components for the retarded dipole
(eqs.~B2-B4 in CRZ2000) are for $t=\rns/\rlc$. Their difference does not
give eqs.~(\ref{dbx} -- \ref{dbz}): 
by overlooking this misalignment of the dipoles
Arendt \& Eilek (1998) incorrectly estimated that the rotational
distortions of the magnetic field are of the order of $\rn$.
For moderate dipole inclinations ($\alpha \sim 40^\circ-50^\circ$) the contours
shown in Fig.~\ref{caps} possess a notch rather than a discontinuous ``glitch"
suggested in Arendt \& Eilek (1998) (see Dyks et al.~2004b for details).}

The following important conclusions can be drawn from Fig.~\ref{caps}:
1) the open volume is strongly asymmetric with respect to the $\om$
plane; 2) the magnitude of the asymmetry depends on $y_{\rm m}$ and thus
on the impact angle $\beta$; 
3) regardless of the value of $\beta$ the pulse window associated with
the outer boundary of the open volume is always shifted
backwards, ie. toward later phases;
4) the magnitude of this rotational asymmetry is very large, 
much larger than the local changes of the direction of the magnetic field
caused by the rotation
(eq.~\ref{kappa}), and even larger than $\rn$ (for large $\alpha$
the ``retarded contours" are (on average)
shifted toward later phases by $\sim 0.2\thov = 0.2
\rn^{1/2} = 0.02$ 
to be compared with $\rn=0.01$).

This numerical result implies that none of the previous estimates
(neither eqs.~\ref{dphasym}, \ref{dphasym2}, nor the Shitov's formula 
\ref{deflection})
provide a reliable measure of the rotational distortion of the \emph{open
volume} shape.
The reason for this is that the boundary of the open volume is not
only determined by the local (ie.~low-altitude) direction of $\vbret$, 
but also (and most importantly) by the geometry of $\vbret$ 
near the light cylinder, where
$\rn\sim 1$. At $\rlc$ all ``higher order" effects become comparable in
magnitude to the lowest order effects,
in the sense that $\rn^m \sim 1$ for any $m$.
With the strength of the rotational distortions being very large at
$\rlc$, a very different set of magnetic field lines is picked up
as the last open field lines which form the boundary of the open volume. 
This ``retarded" boundary is highly asymmetrical with respect to the
$\om$ plane (Fig.~\ref{caps}), and the magnitude of this 
asymmetry has little to do
with the low-altitude rotational distortions as estimated with
eqs.~(\ref{dphasym}), (\ref{dphasym2}) and with Shitov's formula,
because
\emph{the low-altitude crossection of the open volume boundary is
an image of the strong near-$\rlc$ distortions projected 
through the continuity of the magnetic field lines}.
Hereafter, we will refer to the asymmetrical distortion of the open volume 
with the terms ``backward shift" or ``displacement" of open volume.

Another numerical result is that the contours shown in Fig.~\ref{caps}
are (with high accuracy) the same for any radial distance $\rn$, as long
as $\rn \ll 1$, and as long as their size is normalized by $r\thov =
r\rn^{1/2}$, as in Fig.~\ref{caps}. Contours calculated eg.~for $\rn =
10^{-3}$ or $\rn = 0.1$ look exactly the same as those shown 
in Fig.~\ref{caps}. 
This means that for $\rn \ll 1$ the overall (ie.~averaged over the impact
angle $\beta$) angular displacement of the open volume $\displ$
is a fixed fraction 
of the angular radius of the open volume $\thov$:
\begin{equation}
\displ\simeq a \thov \approx a \rn^{1/2}
\label{displ}
\end{equation}
with $a\simeq 0.2\sin\alpha$. The dependence $a\propto\sin\alpha$
has been determined by noting that $\displ$ (measured from the star
center) decreases with $\alpha$ (Fig.~\ref{caps}), 
whereas the corresponding phase shift
$\dphov\approx\displ/\sin\alpha$ of the pulse profile does not 
(cf.~Fig.~\ref{efs}, the discussion in Section 3.1, as well as
eq.~A8 in Dyks et al.~2004a).

Equation (\ref{displ}) implies that the rotational displacement of the open
volume has magnitude comparable to the combined effects of
aberration and propagation time delays 
(hereafter APT effects, of magnitude $2\rn$, cf.~Dyks et al.~2004a) 
for $\rn \sim a^2/4 \sim 0.01$, 
which is quite typical estimate of radio emission
altitudes (eg.~Gupta \& Gangadhara 2003; Kijak \& Gil 2003). For $\rn \ll 0.01$ the backward shift of the
open volume dominates over the APT effects.
For $\rn \gg 0.01$ the APT effects dominate over
the open volume shift.

\subsection{Twofold nature of the rotational distortions}

We find, therefore, that the nature of the rotational distortions
at low altitude is twofold: in addition to the 
famous (but negligible) asymmetrical distortion of the local magnetic field 
direction (of magnitude $\sim \rn^3$, eqs.~\ref{dphasym} and \ref{dphasym2})
the rotation shifts the open volume backward by a much larger amount of
$\sim \rn^{1/2}$. The low-altitude shift of the open volume is 
not caused by the distortions of the shape of magnetic field lines 
\emph{at
low altitudes} --- locally their shapes
are pretty much the same as those of the static dipole. 
It is due to the strong near-$\rlc$ distortions:
magnetic field lines which are picked up as the last open magnetic field lines
by the tangency condition at the light cylinder
are located asymmetrically with respect to the $\om$ plane.
The near symmetry of the field lines at low altitudes 
does not imply similar symmetry of the open volume, because its shape
is
also governed by the geometry of the field near $\rlc$.

Although Shitov (1985) in his analysis of
the phase shift between the pulse profile center and the center of the
position angle curve neglected the APT effects, he did
include the backward displacement of the open volume 
and did emphasize the twofold
nature of the rotational distortions (cf.~his fig.~2).
However, he has not provided any simple estimate of the open volume 
displacement (like eq.~\ref{displ}). Therefore,  
in most studies following his work
only the tiny local deflection of $\vec B$ has been considered
(eqs.~\ref{dphasym}, \ref{dphasym2} and Shitov's estimate),
usually only to infer that the rotational distortions are negligible
in generation of any asymmetry in a pulse profile.

\section{Implications for the relativistic model of pulsar polarization}

Implications of the rotational displacement of the open volume 
for the relativistic model of pulsar polarization are profound, 
because this effect is
\emph{lower} order ($\sim \rn^{1/2}$, eq.~\ref{displ})
than the effects considered so far (aberration $\sim \rn$, propagation time
delays $\sim \rn$).
Although the PA curve is practically unaffected, the center of the
pulse profile to which the PA refers is considerably displaced.\\
Let us define a phase zero as a moment at which an observer 
detects a light signal emitted
from the neutron star center when $\vec \mu$ was in the $\om$ plane.
As discussed in Dyks et al.~2004a, 
the total phase shift $\dphnew$ between the center of the
position angle curve and the pulse profile center can then be separated
into two components: the shift of the center of the PA
curve by $\dphpa \approx 2\rn$ towards later phases 
with respect to the zero phase, 
and the shift of the pulse profile center toward earlier phases by
$\dphp$ with respect to the zero phase.
Had the boundary of the open volume been symmetrical with respect to the
$\om$ plane (as in the case of the static-shape dipole) the profile
center would be shifted forward in phase by $\dphp
\approx -2\rn$ which would result
in the total shift of $4\rn$ as initially predicted by BCW91.
Due to the backward displacement of the open volume given by
eq.~(\ref{displ}), however, the
forward profile shift is decreased by $\dphov \equiv F\rn^{1/2}$
(with $F\sim a/\sin\alpha$) so that $\dphp = -(2\rn - F\rn^{1/2})$.

Therefore, the delay-radius relation of BCW91 (eq.~\ref{bcw}) becomes:
\begin{equation}
\dphnew \equiv \dphpa + (-\dphp) \approx 2\rn + (2\rn-F\rn^{1/2})
\label{dphnew}
\end{equation}
with $F$ in general being a complicated function of $\alpha$, $\zeta$ 
and $\rn$.
The complicated form of $F$ results from the complicated shape of the
open volume boundary (Fig.~\ref{caps}) which implies nontrivial
dependence of $F$ on the impact angle and thereby on $\zeta$.
Since the sign of the impact angle $\beta = \zeta
- \alpha$ provides no information about whether the viewing
trajectory is poleward or equatorward (in the sense defined in Everett
\& Weisberg 2001) hereafter we will use the angle
\begin{equation}
\apan = \left\{
\begin{array}{rl}
  \beta, &\ {\rm if }\ \alpha \le 90^\circ\\
-\beta, &\ {\rm if }\ \alpha > 90^\circ\\
\end{array}
\right.
\label{apan}
\end{equation}
which is negative/positive for poleward/equatorward viewing geometry 
regardless of whether $\alpha > 90^\circ$ or not.
Changes of $F$ as a function of $\tau \equiv
\apan/\thb \approx \apan/(1.5\rn^{1/2})$ are
illustrated in Fig.~\ref{efs} for the same angles $\alpha$ as
those in Fig.~\ref{caps}.
The functions $F(\tau)$ were calculated
for $\rn = 0.01$, however, they change little with $\rn$, as long as
$\rn \ll 1$.
Fig.~\ref{efs} shows that $F$ is confined to the rather limited range 
of $0.1 - 0.4$ for any
combinations of $\alpha$ and $\zeta$. Also, $F$ is always positive which
implies that the displacement of the open volume
results in a smaller phase shift $\dphnew$ than predicted by the original
delay-radius relation (eq.~\ref{bcw}).
The radio emission radii provided by the original delay-radius relation
are therefore underestimated by a factor which may be very large for
some parameters.

\subsection{The misalignment formula}

The delay-radius relation which includes the rotational distortions of
the open volume (in the vacuum approximation) becomes:
\begin{equation}
\dphnew \approx 4\rn - F(\alpha, \zeta, \rn)\ \rn^{1/2}
\label{mf}
\end{equation}
and it 
is plotted in Fig.~\ref{mfig} for a few values of $F$ equal to $0.1$,
$0.2$, $0.3$, and $0.4$ (solid, dotted, dashed, and dot-dashed curve,
respectively). (For convenience, $\dphnew$ on the vertical axis is in
degrees, whereas eq.~\ref{mf} gives $\dphnew$ in radians.)
The thick dashed line presents the original delay-radius relation
of eq.~\ref{bcw}.
For $\rn < F^2/16$ (eg.~for $\rn < 10^{-2}$ for $F=0.4$, dot-dashed line),
ie.~for small emission radii, the phase shift $\dphnew$ becomes negative
(ie.~the center of the PA curve \emph{precedes} in phase the profile
center), which is a new feature in comparison with the original 
delay-radius relation which always predicted the \emph{delay} of the PA
curve with respect to the profile. Since eq.~(\ref{mf}) predicts 
that the center of the PA curve 
may either precede or lag the center 
of the profile, it will be referred to as the ``misalignment" formula.
For large radii $\rn \ga 10^{-2}$ the formula always predicts positive
$\dphnew$ and, for increasing $\rn$, \emph{slowly} converges to the original
delay-radius relation. 

For any $\dphnew$, the radio emission
radii estimated with the original delay-radius relation underestimate those 
implied by eq.~(\ref{mf}) (cf.~Fig.~\ref{mfig}).
For $\dphnew \sim 1^\circ$,
the delay-radius relation underestimates $r$ given by
eq.~(\ref{mf}) by a factor of $1.5-4$ (depending on $F$). 
For $\dphnew \approx 0.3^\circ$, $0.1^\circ$, and $0.01^\circ$ 
the underestimate factor
is in the range of $2- 9$, $3 - 22$, and $16-220$, respectively. 
For $\dphnew \approx 0$ the underestimation factor is formally infinite.
In the absence of effects described in Section \ref{altidep},
the negative values of the shift $\dphnew$ cannot be lower than
$-F^2/16 \ge -0.57^\circ$.  

Unlike the original delay-radius relation,
the misalignment formula predicts that the implied
emission radii depend on the viewing geometry (especially on $\apan$)
and that this dependence cannot be separated from the dependence on $\rn$:
even for fixed $\alpha$ and $\apan$, 
when the line of sight probes the magnetosphere deeper and deeper
(ie.~when $\rn$ decreases in Fig.~\ref{mfig}), the absolute value of 
the parameter $\tau\equiv \apan/(1.5\rn^{1/2})$ increases --- 
the line of sight
cuts through the open volume more peripherally.
This departure from $\tau = 0$ in Fig.~\ref{efs} implies that the value of
$F$ changes (sometimes abruptly) with varying $\rn$, ie.~for fixed
$\alpha$ and $\apan$ the value of $F$ is not fixed --- it depends on $\rn$.

The complicated behaviour of the misalignment formula (\ref{mf}), is
exemplified in Fig.~\ref{mfig2}, which presents $\dphnew$ as a function
of $\rn$ calculated numerically
for $\alpha=45^\circ$ and three values of $\zobs = 43$, $45$, and $47^\circ$
(circles, squares, and crosses, respectively).  
For large $\rn$ (and so for large $\dphnew$)
the line of sight crosses the open volume nearly centrally
($\apan \ll 1.5\thov$, $\tau \simeq 0$), so that $F\simeq 0.1$,
regardless of the value of $\zobs$ and $\alpha$ (cf.~Fig.~\ref{efs}). 
Therefore, all the
three numerical solutions stay close to the analytical solution for
$F=0.1$ (thin solid line). For smaller $\rn$, (and $\dphnew \la
1^\circ$), the numerical results diverge from each other: 
the case of $\zobs=\alpha=45^\circ$ (squares) remains close to the analytical 
solution with $F\simeq 0.1$ (it follows eq.~(\ref{mf}) with $F\approx1.07$),
because the parameter $\tau\propto\apan$ is fixed and equal to zero.\\
In the case of $\zobs=47^\circ$ (equatorward viewing, crosses in
Fig.~\ref{mfig2}), the parameter $\tau$ increases with decreasing $\rn$
because the line of sight traverses more peripherally through the open
volume. As can be inferred from Fig.~\ref{efs} (panel for
$\alpha=45^\circ$) this makes $F$ increase through $0.2$ up to $\sim
0.26$ for $\tau\simeq 1$, and accordingly, the numerical solution in
Fig.~\ref{mfig2} crosses the dotted line for $F=0.2$ and approaches the
vicinity of the dashed line for $F=0.3$.
At $\log \rn \simeq -3.33$ the line of sight just grazes the outer
boundary of the open volume ($\tau \simeq 1.08$). At smaller radial
distances the line of sight does not penetrate the open
volume. \\
In the poleward case of $\zobs=43^\circ$ (circles in Fig.~\ref{mfig2}),
$\tau$ becomes more negative with decreasing $\rn$. Since the backward
displacement of the open volume is stronger on its poleward side
(cf.~Figs.~\ref{caps} and \ref{efs}) the solution crosses the analytical
curve for $F=0.2$ (dotted line in Fig.~\ref{mfig2}) earlier than in the
equatorward case (ie.~at smaller $|\tau|$). At $\log \rn \simeq -2.43$,
the line of sight starts to cut the radiation beam \emph{above} 
the notch visible in Fig.~\ref{caps}. This results in a discontinuous
increase of $F$ from $\sim0.25$ up to $\sim0.39$ (Fig.~\ref{efs}).
Therefore, the numerical solution jumps to the vicinity of the
dot-dashed line of $F=0.4$ (Fig.~\ref{mfig2}). 
For more peripheral traverses (ie.~for smaller $\rn$ and more negative
$\tau$), $F$ changes little between $0.39$ and $0.35$, and the numerical
solution departs only slightly from the $F=0.4$ curve.
For $\log \rn \la -3.25$ the line of sight misses the open volume.
This minimum value of $\log \rn$ differs slightly from the one for
$\apan=+2^\circ$ (crosses) because in addition to the backward
displacement, the open volume is also slightly shifted (with respect to
$\vec \mu$) towards the rotational equator
(ie.~downwards in Fig.~\ref{caps}).

\subsection{Determination of emission radius}
\label{determination}

The above-described complicated form 
of the function $\dphnew(\rn, \alpha, \zobs)$ does not allow
an easy determination of $\rn$. Numerical determination of $\rn$ based
on the known values of $\dphnew$, $\alpha$, and $\zobs$ requires 
rather complicated calculations. Therefore, below we 
discuss the particular cases when easy derivation of $\rn$ is possible
and then we propose
a procedure, which allows the possible range of $\rn$
to be constrained
in a general case of arbitrary viewing geometry.

\begin{figure}
\epsscale{1.0}
\plotone{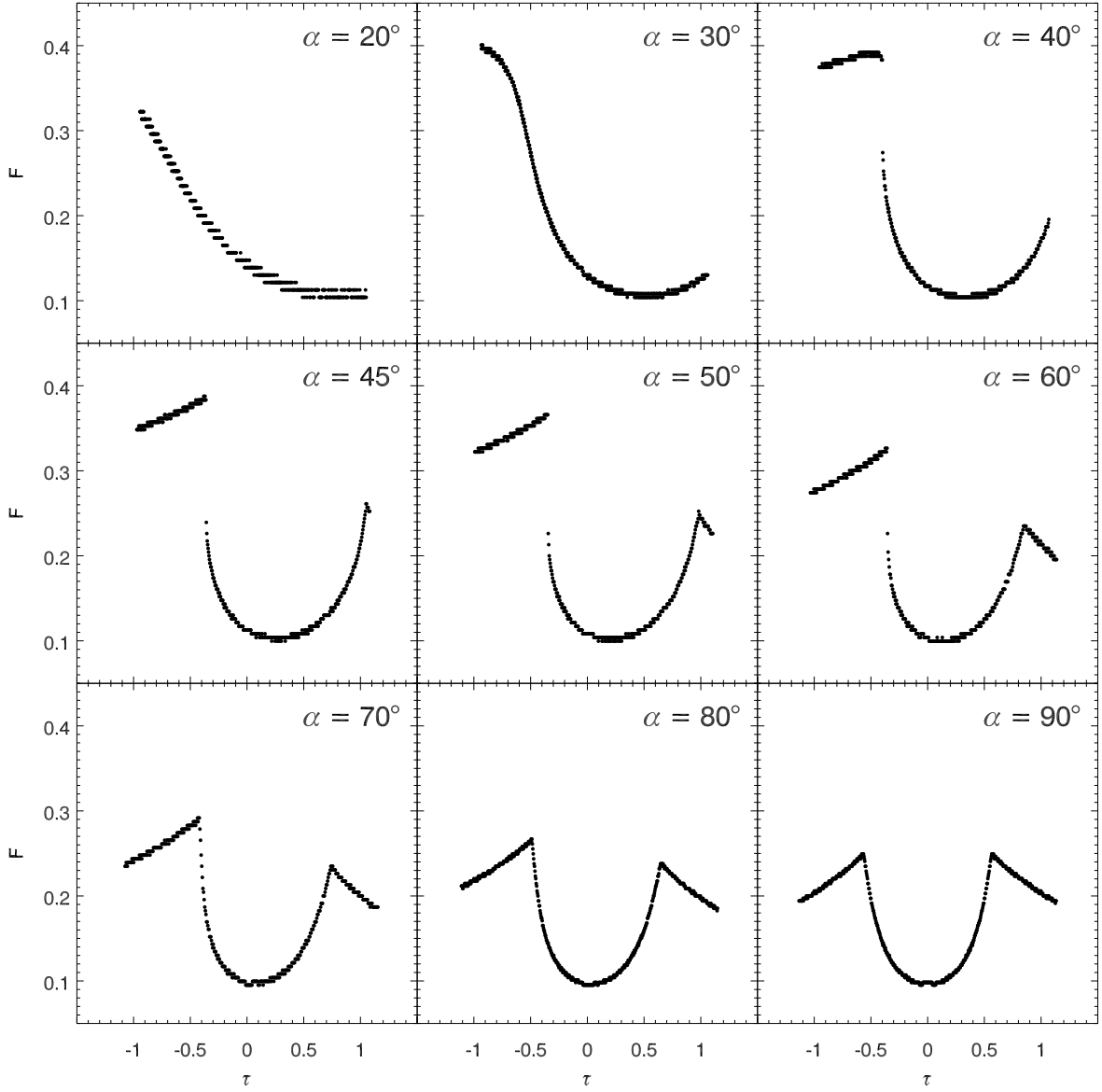}
\caption{The parameter $F$ of eq.~\ref{dphnew} as a function of
the parameter $\tau = \apan/(1.5\rn^{1/2})$ for the same dipole
inclinations $\alpha$ as in Fig.~\ref{caps}. Though the results were
obtained for $\rn=0.01$, the curves change little with $\rn$, as long as
$\rn \ll 1$.  
\label{efs}}
\end{figure}

\begin{figure}
\epsscale{0.7}
\plotone{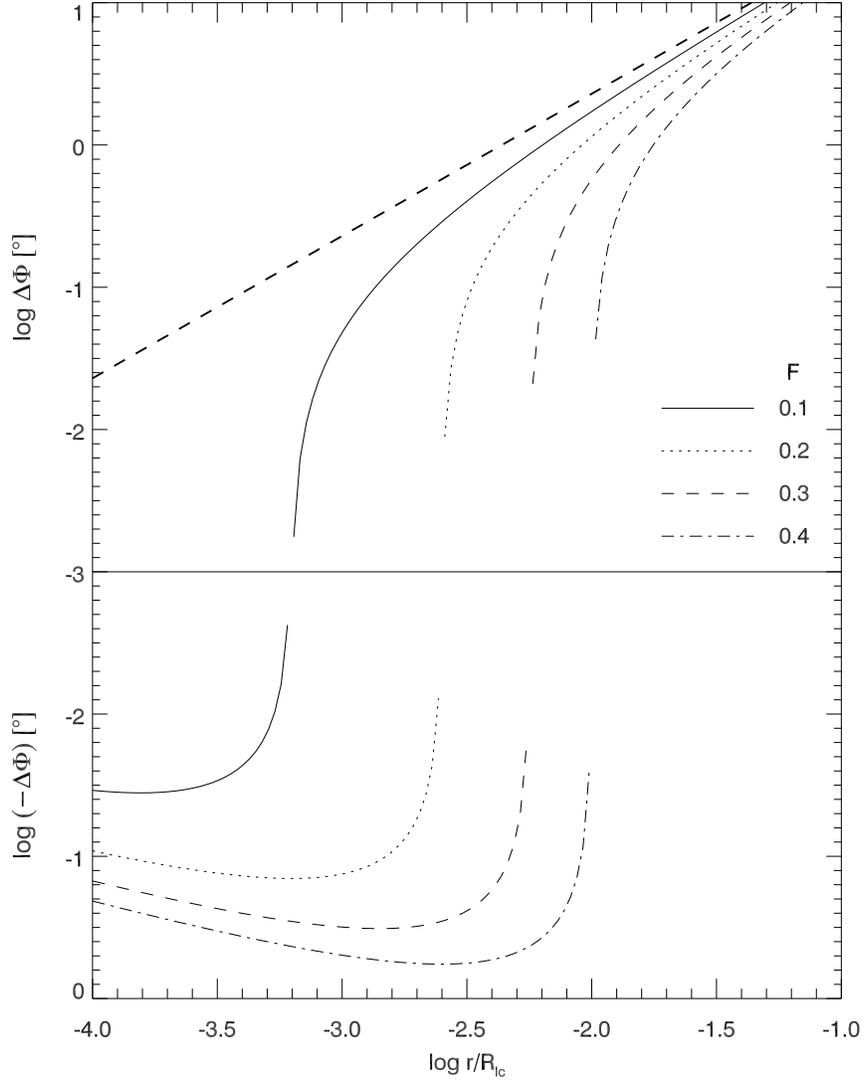}
\caption{The phase shift between the center of the PA curve and the
profile center as a function of the radial distance of the radio
emission. The four curves show the case with the rotational distortion
of the open volume included (eq.~\ref{mf}) 
and correspond to different values of the
parameter $F$ shown in the lower right corner of the upper panel. 
The thick dashed line presents the original
delay-radius relation (eq.~\ref{bcw}) which does not include
the sweepback effect. In the upper panel the positive shift is shown
(the PA curve lags the profile). The lower panel is for the negative
shift (the PA curve precedes the pulse profile). 
Note that the original delay-radius relation significantly
underestimates $\rn$, especially for small $\dphnew$ and $r/\rlc$.
\label{mfig}}
\end{figure}

\begin{figure}
\epsscale{0.7}
\plotone{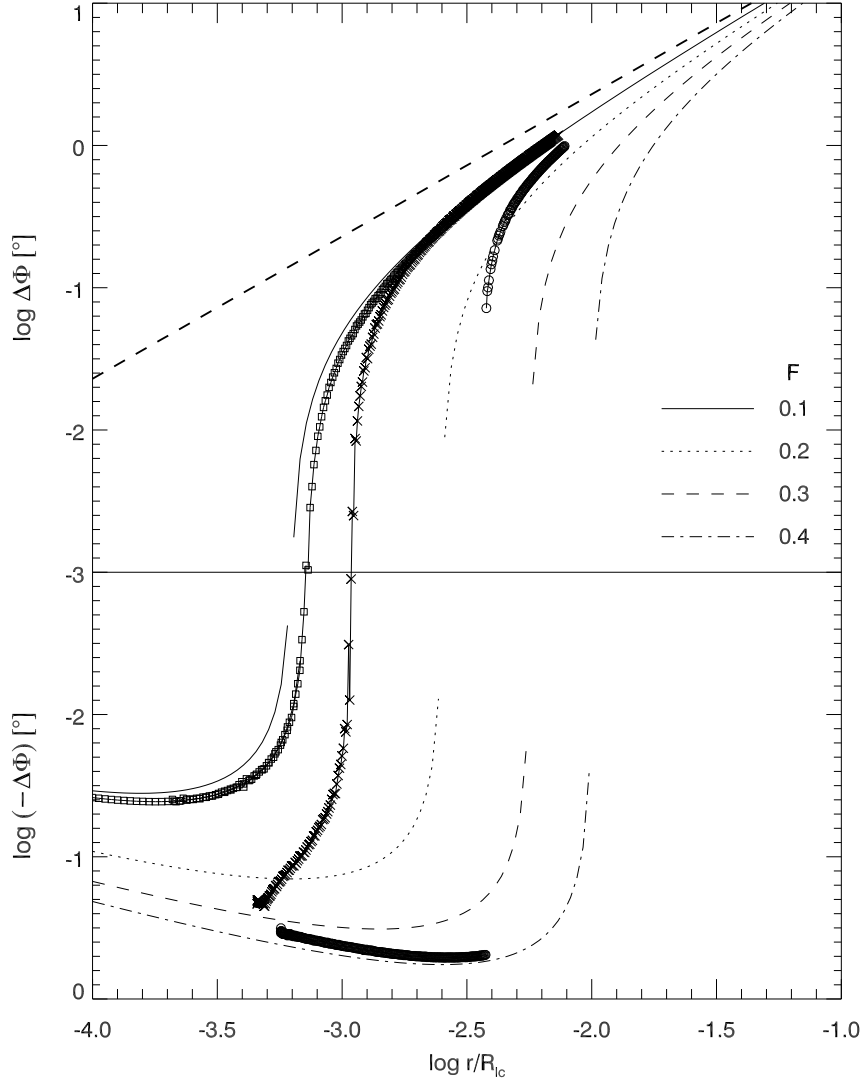}
\caption{Three numerical results for $\alpha=45^\circ$
(and different viewing angles $\zeta$) overplotted on the curves from 
Fig.~\ref{mfig}. The circles are for the poleward viewing geometry
with $\zeta=43^\circ$ (the curve
is broken into two parts, one with positive, and the other with negative 
$\dphnew$), 
the crosses are for the equatorward viewing with $\zeta = 47^\circ$,
and the squares are for $\alpha=\zeta=45^\circ$.
For more details see text. 
\label{mfig2}}
\end{figure}

As can be inferred from Fig.~\ref{efs}, the value of $F$ is (nearly) fixed
and close to $0.1$ whenever $|\tau| \ll 0.4$, 
ie.~whenever $|\beta| \ll 0.6\rn^{1/2}$.
The condition is fulfilled 
for any $\rn$ and $\alpha$
when $\beta\approx 0$ ($\alpha\approx\zobs$,
squares in Fig.~\ref{mfig2}). 
The other case when the condition is fulfilled is 
when the measured shift between 
the position angle center and the profile center
is large ($\ga 1^\circ$) and the impact angle $|\beta|$ is small: 
the radiation comes then from high-altitudes,
where $\thov \gg |\beta|$.
This is why all numerical results shown in Fig.~\ref{mfig2} approach the
solid line of $F=0.1$ when $\dphnew$ increases above $1^\circ$.

When $F$ is fixed (like in the above-described cases), 
the equation (\ref{mf}) can be inverted
to obtain analytical solutions for $\rn(\dphnew)$:
\begin{equation}
\rn = 32^{-1}\left[8\dphnew + F^2 + 
k\left(16\dphnew F^2 + F^4\right)^{1/2}\right] 
\label{r}
\end{equation}
where the parameter $k=1$ for $\dphnew > 0$ whereas $k=\pm1$ for
$\dphnew < 0$. Thus, the solution for $\rn$ is unique if $\dphnew$ 
is positive, whereas 
for the negative $\dphnew$ \emph{two} solutions are possible.
One can attempt to reject the smaller one of these solutions 
by using the theoretical constraints: 
\begin{equation}
\rn \ge \frac{\rns}{\rlc}\ \ {\rm and}\ \ \rn \ge \frac{4}{9}\ \beta^2.
\label{constr}
\end{equation}
 The second constraint
holds when the emission is limited to the open volume.\footnote{
In application to
the retarded dipole field this condition is approximate, not exact.
Therefore, the minimum values of $\log \rn$ determined numerically 
for $\beta=\pm2^\circ$ in Fig.~\ref{mfig2} (circles and crosses)
differ slightly from $-3.266$.}
In the two above-mentioned cases (when $\beta\approx 0$ or 
$\dphnew \ga 1^\circ$), one
can directly use eq.~(\ref{r}) with $F=0.1$ to calculate $\rn$.

In a general case (including the common case when $\beta\ne 0$ 
and the shift is small) one can
constrain $\rn$ to a degree which depends on the information available.
Without the knowledge of $\alpha$ and $\beta$ one
can obtain a \emph{rough} estimate of $\rn$ based on the measured shift 
$\dphnew$ by assuming some averaged, fixed value of $F$ in eq.~\ref{mf}
(eg.~$F = 0.2$). This is equivalent to igoring the dependence of
$F(\alpha, \tau)$ shown in Fig.~\ref{efs}. For $\rn \le 10^{-2}$ 
this may give results wrong by as much as an order of magnitude
due to the large horizontal spread of $\rn$ as allowed by the range of
$F$ (Fig.~\ref{mfig}).\\
Therefore, a better approach is to use the following two-step procedure:
1) Using eq.~(\ref{r}) one can calculate the
\emph{range} of $\rn$ allowed by $0.1\le F \le 0.4$ and by the
conditions (\ref{constr}). Or, one can draw a horizontal line in 
Fig.~\ref{mfig} to determine graphically the range of $\rn$ 
allowed by the range of $F$ (the measurement error of
$\dphnew$ can easily be taken into account by drawing a horizontal 
strip instead of the line).
For $-0.0358^\circ \le \dphnew < 0$, two
ranges of $\rn$ are allowed and again the constraints
(\ref{constr}) may be used to try to reject the lower range.
For $-0.573^\circ \le \dphnew < -0.0358^\circ$, the allowed range of
$\rn$ is between $\rn^{\rm min}$ and  $\rn^{\rm max}$
where $\rmin$ is given by eq.~(\ref{r}) with $k=-1$ and $F=0.4$
and $\rmax$ by eq.~(\ref{r}) with $k=+1$ and the same value of
$F=0.4$.\\
2) With the knowledge of $\alpha$ and $\beta$ one can more tightly
constrain the allowed range of $\rn$ by calculating $\tmin=\tau(\rmax)$ 
and $\tmax=\tau(\rmin)$ where $\rmin$ and $\rmax$ were determined in
the previous step. Then, from Fig.~\ref{efs}, using the appropriate panel, 
one can determine the narrower range of $(\fmin, \fmax)$ corresponding to the
range of $(\tmin, \tmax)$. Then, one returns to the step 2) in which
the tightened range $(\fmin, \fmax)$ must be used instead 
of the original range of $(0.1, 0.4)$.

If the sign of $\beta$ is unknown, the step 2) may (but does not have
to) give two ranges of $\rn$. If the sign is positive (equatorward
viewing) the range of $F$ considered in the step 1) may be narrowed
to $0.1 - 0.25$ (Fig.~\ref{efs}).

As an example we apply the method to PSR B0301$+$19 and B0525$+$21.
In the case of B0301$+$19 ($P=1.38$ s), BCW91 find $\dphnew =
0.2\pm0.1^\circ$ at $1.4$ GHz. The range of $\rn$ for $F$ between $0.1$ 
and $0.4$ is $(\rmin,\rmax)=(2.0\cdot10^{-3}, 1.2\cdot10^{-2})$,
or, if we allow for the error of $\dphnew$, the range becomes 
$(1.4\cdot10^{-3}, 1.25\cdot10^{-2})$ (step 1).
For $\apan=-0.96\pm0.63^\circ$ (Everett \& Weisberg 2001, hereafter EW2001) 
Fig.~\ref{efs} gives $(\tmin, \tmax)=(-0.3, -0.1)$ 
(with the error of $\dphnew$
included) or $(\tmin, \tmax)=(-0.5, -0.03)$ (including also the error of
$\apan$). EW2001 derived $\alpha = 162.4\pm11.8^\circ$, so that
$\pi-\alpha \approx 17.6^\circ$ and we may use the panel of
Fig.~\ref{efs} for $\alpha=20^\circ$ to constrain the range of $F$
to $(\fmin,\fmax)=(0.14, 0.22)$.
This new range of $F$ translates to $(\rmin, \rmax)=(2.7\cdot10^{-3},
4.6\cdot 10^{-3})$ if the error of $\dphnew$ is neglected 
(and the error of $\apan$ is included). Taking into account
the error of $\dphnew$ one finally obtains $(\rmin, \rmax)=(2.0\cdot10^{-3},
5.3\cdot10^{-3})$. None of the constraints (\ref{constr}) narrows this
range.
This result is in agreement with the condition $\rn \ge \rgeo/\rlc$.
For the observed pulse width $W\approx 15.9^\circ$
(BCW91) and for $\alpha$ and $\beta$ cited above, one finds $\rgeo/\rlc
\approx 1.3\cdot10^{-3}$. Our value of
$\rn=(2-5.3)\cdot10^{-3}$ implies that 
the emission associated with the profile edge
must originate from magnetic field lines with magnetic colatitude
$\thm \approx (0.5-0.8)\thov$. The original delay-radius relation
gives $\rn \approx 9\cdot10^{-4}$ (3 times smaller) which results in 
$\thm/\thov > 1$ (emission from the region of closed field lines).

In the case of B0525$+$21 ($P=3.74$ s), 
BCW91 find $\dphnew=0.3\pm0.1^\circ$ at $430$
MHz. At this value of $\dphnew$ the range of $F=0.1-0.4$
translates into $(\rmin,\rmax)=(2.6\cdot10^{-3}, 1.2\cdot10^{-2})$
or into $(2\cdot10^{-3}, 1.3\cdot10^{-2})$ if the error of $\dphnew$
is included. For $\apan = 1.5\pm0.08^\circ$ (EW2001) we find $(\tmin,
\tmax) = (0.15,0.39)$. EW2001 suggest $\pi-\alpha = 63.2^\circ$ which
implies that $F\simeq0.1$ (Fig.~\ref{efs}, panel for
$\alpha=60^\circ$). Using this value in eq.~(\ref{mf}) one finds that
$\rn=(2.6\pm0.6)\cdot10^{-3}$ which is two times larger than
the value predicted by the original delay-radius relation
(eq.~\ref{bcw}).  
However, the value is still six times smaller than
$\rgeo/\rlc=1.7\cdot10^{-2}$ (calculated for $W=20.4^\circ$) and implies that
the emission at the profile edge comes from magnetic field lines with
colatitudes $\thm =(2.4-3.0)\thov$, ie.~from the closed line region.
A large part of this discrepancy can be removed by considering different
emission altitudes across the pulse profile.  

\section{Altitude-dependent position angle swing}
\label{altidep}

\begin{figure}
\epsscale{0.7}
\plotone{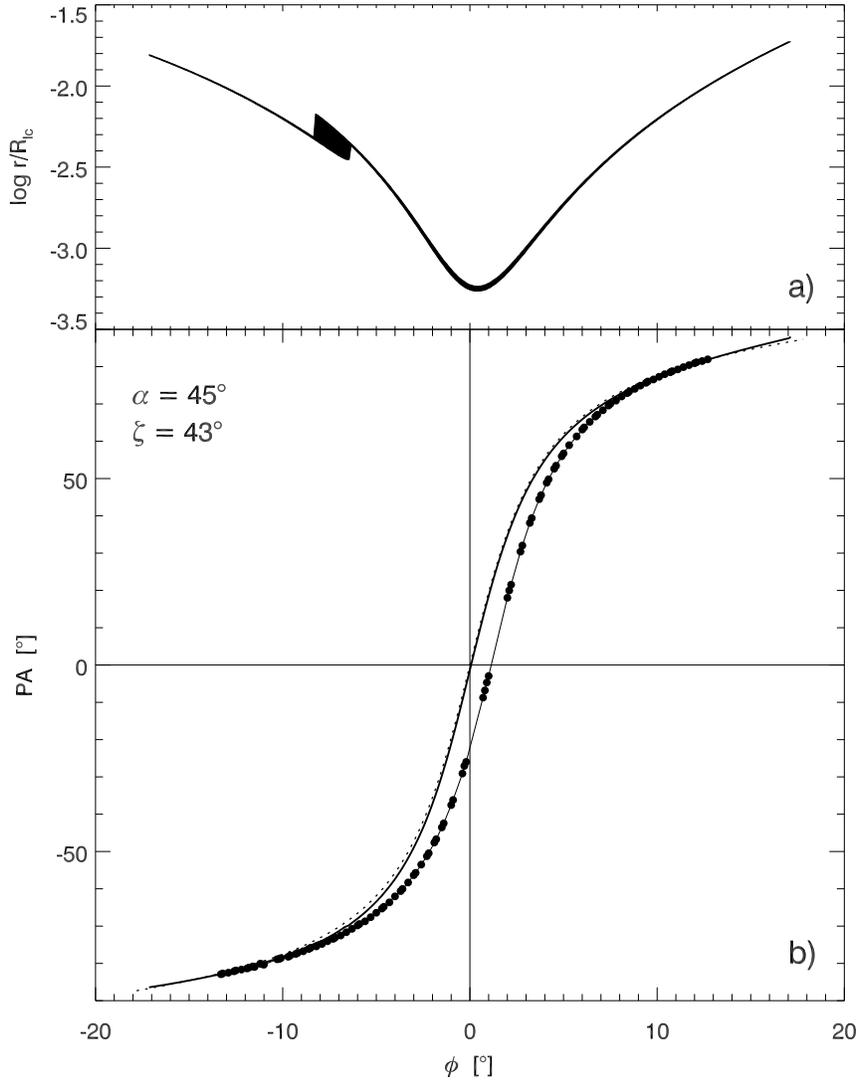}
\caption{Influence of variations of emission altitudes across the pulse
profile on the shape of the PA curve. Thin solid line with dots (panel b) 
presents the position angle curve
for emission from the fixed radial distance
of $\rn =0.01$. 
Its center lags the phase zero by $2\rn\ {\rm rad} \approx 1.14^\circ$.
The thick solid line in panel b is for the radio emission
from the last open field lines, ie.~it corresponds to different
radial distances shown in panel a. Note that the PA curve for the case
of varying $r$ does not exhibit any noticeable lag. The spread in $r$
visible in panel a) near $\phi = -7^\circ$ corresponds to the notch in the
open volume which appears for moderate dipole inclinations
(cf.~Fig.~\ref{caps}, $\alpha=45^\circ$). The dotted line (nearly
overlapping with the thick solid) is the curve of
Radhakrishnan \& Cooke (1969), undisturbed by the special relativistic
effects.
\label{lowcore}}
\end{figure}

When different parts of the pulse profile originate from different
altitudes, the PA curve can no longer be described by the
standard equation of 
Radhakrishnan \& Cooke (1969). The simple analytical equation for the
altitude-dependent PA swing is given in eq.~(16) of BCW91.
The formula predicts that if the radial distance $r$ of the radio
emission is uniform across the pulse profile, 
the entire PA curve is shifted rightwards
(toward later phases) by $2r$ with respect to the zero phase. 
This effect is illustrated in Fig.~\ref{lowcore}: The thin solid line
with dots is the PA curve for the fixed emission altitude of
$\rn=10^{-2}$, calculated numerically for the retarded dipole field
with $\alpha=45^\circ$ and $\zeta=43^\circ$.
In agreement with BCW91, its center lags the phase zero by $2\rn\ {\rm rad} = 
1.14^\circ$,
and there is no discernible sign of the sweepback effect
(see Section 2).

If the central parts of the pulse profile originate from much lower
radial distance $r$ than the edge, 
and if $r \ll 0.01\rlc$,
the central parts of the PA curve do not exhibit
any appreciable shift and nearly follow the undisturbed S-curve of
Radhakrishnan \& Cooke.
The solid line which nearly passes through the $(0,0)$ point in
Fig.~\ref{lowcore}b has been calculated for the emission from the last
open magnetic field lines of the retarded dipole
for the same $\alpha$ and $\zeta$ as above. 
The corresponding radial distance of
the radio emission as a function of phase $\phi$ is shown in panel a of
Fig.~\ref{lowcore}. Within the central parts of the PA curve, the
emission altitude is negligibly small and
the PA curve nearly follows the undisturbed S-swing
(dotted line in Fig.~\ref{lowcore}b).

\begin{figure}
\epsscale{0.7}
\plotone{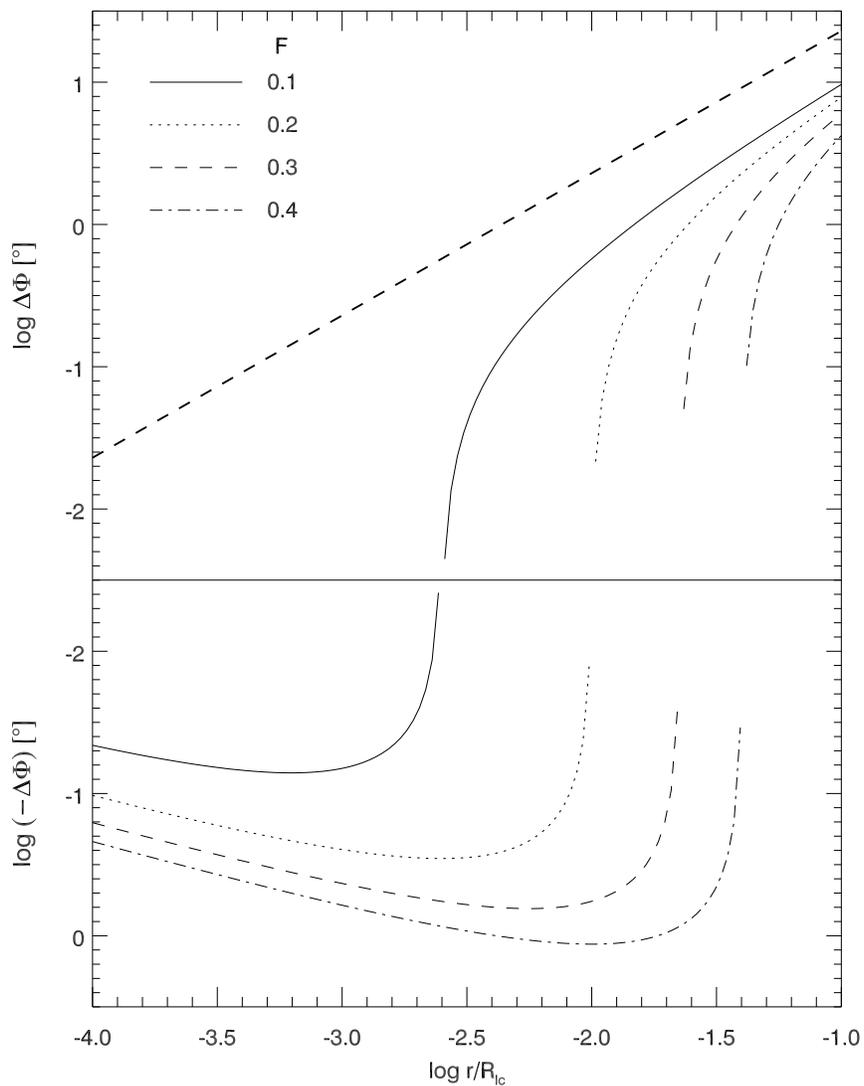}
\caption{Dependence of the shift between the center of the PA curve
and the profile center on the radial distance of the radio emission
in the case when the central parts of the pulse profile originate from
much lower altitude than its edge (ie.~for the case marked in
Fig.~\ref{lowcore} with the thick solid line). The layout is the same as
in Fig.~\ref{mfig}. Note the increased divergence from the original
delay-radius relation (thick dashed line). 
\label{mfig3}}
\end{figure}

Thus, in the case when the central parts of the profile are emitted from much
lower
altitudes than the outer edge of the profile, the PA swing practically
does not undergo the delay by $2\rn$. The misalignment formula
(\ref{mf}) then becomes
\begin{equation}
\dphnew \approx 2\rn - F\rn^{1/2}
\label{mf2}
\end{equation}
and is illustrated in Fig.~\ref{mfig3} with the same layout as
Fig.~\ref{mfig}.
One can see that in the considered case, 
the shift is much smaller than the shift predicted by the standard delay-radius 
formula (eq.~\ref{bcw}, thick dashed line in Fig.~\ref{mfig3}).
For a typical phase shift $\dphnew \sim 0.3^\circ$ eq.~(\ref{bcw})
underestimates $r$ by an order of magnitude.
The range of altitudes for which the shift is negative is much larger
than in the case with the constant emission altitude
(eq.~\ref{mf}, Fig.~\ref{mfig}).

In the considered case of the low-altitude emission within the profile
center, the determination of the radial distance of radio emission $\rn$
is performed as before (Section \ref{determination}) with the only
difference in that eq.~(\ref{r}) should be replaced by the inverse of
eq.~(\ref{mf2}), ie.~Fig.~\ref{mfig3} should be used instead of
Fig.~\ref{mfig}.
Even allowing for the low-altitude origin of the central parts of the
pulse profile, for B0525$+$21 at 430 MHz we still have $\rn < \rgeo/\rlc$.
Apparently, either the radio beam of this pulsar does not fill in
the open volume or other effects are important.
These may include the overestimate of dipole inclination and/or 
$|\beta|$, the broadening
of the observed pulse width due to the low energy of radio emitting
electrons (all of which lead to the overestimate
of $\rgeo$) or refraction, eg.~Lyubarski \& Petrova (1998). 
However, our radial distance $\rn \simeq 7\cdot 10^{-3}$ underestimates
$\rgeo/\rlc$ by a factor of $2.5$ whereas the original delay-radius
formula underestimates it by a much larger factor of $13$.

\section{Conclusions}

The rotational distortions of the vaccum dipole have twofold effect:
in addition to the small changes of the local direction of the magnetic
field, the region of the open magnetic field lines 
undergoes a strong distortion.

The change of the local direction of $\vec B$
is second order in $\rn$. We find, however, that it is symmetrical
with respect to the $\om$ plane. The largest
asymmetrical change of $\vec B$ direction is much smaller -- 
third order in $\rn$, in agreement with Shitov's estimate.

The displacement of the open volume shifts the pulse window toward later
phases with respect to the center of the position angle curve.
The shift has the magnitude of the order of $\rn^{1/2}$.
The open volume shift modifies the delay-radius relation
if the center of the pulse profile is determined as the midpoint between
the outer edges of the pulse profile, which are assumed to lie close to 
the outer boundary of the open volume.
At low altitudes, where effects of aberration and propagation time
delays are small, the open volume shift dominates and may result 
in the center of the PA curve \emph{preceding} the center of the
profile (negative $\dphnew$). A majority of pulsars exhibits
positive $\dphnew$ which means that the radio emission altitudes
typically exceed $\sim 10^{-2}\rlc$. 

The radii derived with the misalignment formula exceed those derived
with the original delay-radius relation by a factor which increases
quickly with decreasing altitude (and $\dphnew$). This explains the trend 
of the delay-radius relation to predict emission radii smaller than
the geometrical radii. 
The underestimate may also be produced/enhanced
by the low-altitude emission within the central parts of the pulse profile.
When both these effects work together, the standard delay-radius
relation may underestimate $r$ by an order of magnitude
even for the relatively large phase shifts ($\dphnew \sim 0.3^\circ$).
 
The influence of the open volume shift on the method
based on the core-cone shift (eq.~\ref{drh}) is difficult to assess,
because the locations of the conal maxima do not need to follow the open
boundary of the open volume. If they did, eq.~(\ref{drh}) would have to
be replaced with $\Delta\phi_{\sss DRH} \approx 2\rn - F\rn^{1/2}$.
The influence of the open volume shift on the geometrical method
is small because the rotation increases the transverse size 
of the open volume insignificantly (by a factor smaller than $\sim 1.2$,
cf.~Fig.~\ref{caps}).

The discussed distortion of the open volume is generated by a
high-order effect which was being neglected for years on the basis
that its ``local magnitude" at low altitudes is small ($\sim \rn^2$).
The actual importance of this effect appears to be much larger than
that. This suggests that other ``high-order" effects,
eg.~the longitudinal polar cap currents, of magnitude $\sim \rn^{3/2}$,
or the inertia of electrons, may be a lot more significant
than their low altitude magnitude suggests. 
The toroidal currents due to the corotation of the charge-filled
magnetosphere ($\sim \rn^2$) have been shown to notably
modify the shape of the open volume, however, in a way which is
symmetrical with respect to the $\om$ plane (cf.~fig.~4.11 
in Beskin et al.~1993).

\acknowledgments
  
We thank U.~Dyks for the derivation of eq.~(\ref{kappa}).
JD thanks B.~Rudak for all the years of fruitful collaboration.
This work was performed while JD held a National Research
Council Research Associateship Award at NASA/GSFC.
This work was also supported by the grant ... (JD) 
and by the NASA Astrophysics Theory Program (AH).

\appendix
\section{Retarded vs static dipole}
\label{appenA}

In the reference frame with $\hat z \parallel \vec \Omega$
the ``retarded" field $\vbret$ 
of the vacuum magnetic point dipole with the magnetic
moment given by eq.~(\ref{mu}) has the following cartesian components:
\begin{eqnarray}
\bretx & = & \frac{\mu}{r^5}
\left\{\vphantom{\frac{r}{r}}3xz\cos\alpha\ + \right. \sin\alpha\left( \vphantom{\frac{r}{r}}\right.
 \left[(3x^2 - r^2) + 3xy\rn + (r^2 - x^2)\rn^2\right] 
\cos(\Omega t - \rn)+\nonumber\\
        &    &
\left[3xy - (3x^2 - r^2)\rn - xy\rn^2\right] \sin(\Omega t - \rn)
\left.\left.\vphantom{\frac{r}{r}}\right)\right\}
\label {bretx} \\
\brety & = & \frac{\mu}{r^5}
\left\{\vphantom{\frac{r}{r}}3yz\cos\alpha\ + \right. \sin\alpha\left( \vphantom{\frac{r}{r}}\right.
 \left[3xy + (3y^2 - r^2)\rn - xy\rn^2\right] 
\cos(\Omega t - \rn)+\nonumber\\
        &    &
\left[(3y^2 - r^2) - 3xy\rn +(r^2 - y^2)\rn^2\right] \sin(\Omega t - \rn)
\left.\left.\vphantom{\frac{r}{r}}\right)\right\}
\label {brety} \\
\bretz & = & \frac{\mu}{r^5}
\left\{\vphantom{\frac{r}{r}}(3z^2 - r^2)\cos\alpha\ + \right. \sin\alpha\left( \vphantom{\frac{r}{r}}\right.
 \left[3xz + 3yz\rn - xz\rn^2\right] 
\cos(\Omega t - \rn)+\nonumber\\
        &    &
\left[3yz - 3xz\rn - yz\rn^2\right] \sin(\Omega t - \rn)
\left.\left.\vphantom{\frac{r}{r}}\right)\right\}
\label {bretz}
\end{eqnarray}
It can be derived by any of the methods described in Yadigaroglu (1997),
Arendt \& Eilek (1998), or CRZ2000. Also, it can be obtained by taking
the limit of $\rns/\rlc \ll 1$ in the solution of Deutsch (1955).
For any position $\vec r_0$, time $t_0$ and time interval $\Delta t$
it holds that 
\begin{equation}
\vbret\left(t_0+\Delta t, R_z(\Omega \Delta t)\vec r_0\right) =
R_z(\Omega \Delta t) \vbret\left(t_0, \vec r_0\right), 
\label{rotation}
\end{equation}
where $R_z(\omega)\vec r$
represents the rotation of the vector $\vec r$ by the angle $\omega$
around the $\hat z$ axis. 
Thus, the dependence on the time $t$ only reflects the rigid rotation
of the pattern of distorted magnetic field lines
around the $\hat z$ axis with the angular velocity $\Omega \hat z$.
The magnetic field $\vbstat$ of 
the static-shape dipole associated with the same magnetic moment 
$\vec \mu$ (eq.~\ref{mu})
is given by the same set of equations
(\ref{bretx} -- \ref{bretz}) with $\rn=0$.

Using eqs.~(\ref{bretx} -- \ref{bretz}) with $t=0$ 
it can be immediately shown that the
difference $\vdb = \vbret - \vbstat$ is given by eqs.~(\ref{dbx} --
\ref{dbz}). Using (\ref{db}) and (\ref{dbx} -- \ref{dbz}) one can find
that the angle $\kappa=\arccos(\vbret\cdot\vbstat/(\bret\bstat))$
is given by eq.~(\ref{kappa}) with $f_1$, $f_2$, and $f_3$ given by 
\begin{equation}
f_1 = h_1 - g_1^2,\ \ \ \ \ f_2 = h_2 - 2 g_1 g_2 
\label{smallefs}
\end{equation}
and 
\begin{equation}
f_3 = 2 g_1^3 + h_3 - g_2^2 - 2 g_1 (g_3 + h_1)
\label{smallefs2}
\end{equation}
with
\begin{eqnarray}
g_1 & = & \frac{1}{2}\ c_1 \left[\left(5\frac{x^2}{r^2} - 1\right)
\sin\alpha + 5\frac{x^2}{r^2}\cos\alpha\right]\\
g_2 & = & -2 c_1 \left[
\frac{xy}{r^2}\sin\alpha + \frac{y^2}{r^2}\cos\alpha\right]\\
g_3 & = & \frac{1}{8}\ c_1 \left[
\left(3 - 7\frac{x^2}{r^2}\right)\sin\alpha - 7\frac{x^2}{r^2}\cos\alpha\right]\\
h_1 & = & \frac{1}{4}\ c_1 \sin\alpha \left(3\frac{x^2}{r^2} + 1\right)\\
h_2 & = & -\frac{2}{3}\ c_1 \sin\alpha \frac{x}{r}\frac{y}{r}\\
h_3 & = & -\frac{1}{8}\ c_1 \sin\alpha\left(\frac{x^2}{r^2} -
\frac{5}{9}\right)\\
c_1 & = & \frac{\sin\alpha}{1 + 3 \left(xr^{-1}\sin\alpha +
zr^{-1}\cos\alpha\right)^2}.
\end{eqnarray}
The function $f_4$ in eq.~(\ref{dphasym2}) is equal to
\begin{equation}
f_4 = \frac{1}{2}\left(\frac{\mu}{r^3}\right)^2 
\frac{\sin\alpha}{\bstatx^2}\ 
\frac{y}{r}
\left(3\frac{z}{r}\cos\alpha + 4 \frac{x}{r}\sin\alpha\right),
\label{ef4}
\end{equation}
where $\bstatx$ is given in eq.~(\ref{bstatx}).

The cartesian components of the retarded magnetic field given by 
eqs.~(\ref{bretx} -- \ref{bretz}) can be rewritten into the following
spherical components of $\vbret$ in the reference frame
with $\hat z \parallel \vec \Omega$: 
\begin{eqnarray}
\bretr  & = & \frac{2\mu}{r^3} \left\{ 
\cos\alpha\cos\theta + \sin\alpha\sin\theta \left[
  \rn \sin\lambda + \cos\lambda \right] \right\}
\label {bret1} \\
\bretp & = & -\frac{\mu}{r^3} \sin\alpha \left[
\left(\rn^2 - 1 \right) \sin\lambda 
+ \rn \cos\lambda \right], 
\label{bret2} \\
\brett & = & \frac{\mu}{r^3} \left\{ 
\cos\alpha\sin\theta + \sin\alpha\cos\theta \left[ 
- \rn \sin\lambda + \left(
\rn^2 - 1 \right)
\cos\lambda \right] \right\}
\label{bret3}
\end{eqnarray}
where
\begin{equation}
\lambda = \rn + \phi - \Omega t.
\label{lambda}
\end{equation}
Again, the (spherical) components of $\vbstat$ are given by 
eqs.~(\ref{bret1} -- \ref{lambda}) with $\rn = 0$.   

Using eqs.~(\ref{bret1} -- \ref{bret3}) in eq.~(\ref{db})
it can be easily shown that
\begin{eqnarray}
\dbr & = & \frac{\mu}{r^3} \sin\alpha\sin\theta 
\left[ \rn^2 \cos\delta
-\frac{2}{3}\ \rn^3 \sin\delta
+ O\left(\rn^4\right)\right] \\
\dbp & = & \frac{\mu}{r^3} \sin\alpha
\left[-\frac{1}{2}\ \rn^2 \sin\delta   -\frac{2}{3}\ \rn^3 \cos\delta
+ O\left(\rn^4\right)\right]\\
\dbt & = & \frac{\mu}{r^3} \sin\alpha\cos\theta 
\left[ \frac{1}{2}\ \rn^2 \cos\delta
-\frac{2}{3}\ \rn^3 \sin\delta
+ O\left(\rn^4\right)\right],
\end{eqnarray}
where $\delta = \phi - \Omega t$.

\clearpage

\end{document}